\documentclass{emulateapj}

\begin{document}

\shortauthors{Luhman \& Esplin}

\shorttitle{Spectroscopy of Sco-Cen Candidates}

\title{Spectroscopy of Candidate Members of the Sco-Cen Complex\altaffilmark{1}}

\author{K. L. Luhman\altaffilmark{2,3}
and T. L. Esplin\altaffilmark{4}}

\altaffiltext{1}
{Based on observations made with the Gaia mission, the Two Micron All
Sky Survey, the NASA Infrared Telescope Facility,
Cerro Tololo Inter-American Observatory, Gemini Observatory, and
the European Southern Observatory.}

\altaffiltext{2}{Department of Astronomy and Astrophysics,
The Pennsylvania State University, University Park, PA 16802, USA;
kll207@psu.edu}

\altaffiltext{3}{Center for Exoplanets and Habitable Worlds, The
Pennsylvania State University, University Park, PA 16802, USA}

\altaffiltext{4}{Steward Observatory, University of Arizona, 933 North Cherry
Avenue, Tucson, AZ 85721, USA}

\begin{abstract}

We present spectroscopy of 285 previously identified candidate
members of populations in the Sco-Cen complex, primarily Ophiuchus,
Upper Sco, and Lupus. The spectra are used to measure spectral types
and diagnostics of youth. 
We find that 269 candidates exhibit signatures of youth in our spectra
or previous data, which is consistent with their membership in Sco-Cen.
We have constructed compilations of candidate members of Ophiuchus, Upper Sco,
and Lupus that have spectral classifications and evidence of youth,
which contain a total of 2274 objects.
In addition, we have used spectra from previous studies to
classify three sources in Ophiuchus that have been proposed
to be protostellar brown dwarfs: ISO~Oph~70, 200, and 203. 
We measure spectral types of early M from those data, which
are earlier than expected for young brown dwarfs based on
evolutionary models ($\gtrsim$M6.5) and instead are indicative of
stellar masses ($\sim0.6$~$M_\odot$).

\end{abstract}

\section{Introduction}
\label{sec:intro}

Spectroscopic surveys in nearby star-forming regions are important
for providing large samples of members that have measured spectral types
and confirmation of youth, which can serve as the foundation for
various studies of star and planet formation \citep{hil97}.
The populations within the Scorpius-Centaurus complex \citep[Sco-Cen,][]{pm08}
are appealing targets for such surveys because of their richness and proximity
\citep[100--200~pc,][]{dez99}.
Those populations consist of the stars associated with the Ophiuchus and
Lupus clouds, a compact group surrounding the star V1062~Sco, Upper Scorpius,
Upper Centaurus-Lupus (UCL), and Lower Centaurus-Crux (LCC), where the
latter two are now known to comprise a single continuous distribution of
stars \citep{riz11,luh21}.

A variety of methods have been used to identify
candidate members of Sco-Cen \citep{com08,pm08,wil08}.
Recent progress has been enabled by the high-precision astrometry 
from the Gaia mission \citep{per01,deb12,gaia16b,bro18,bro21}
and optical and infrared (IR) photometry from wide-field surveys, including
the Two Micron All Sky Survey \citep[2MASS,][]{skr06}, 
the United Kingdom Infrared Telescope Infrared Deep Sky Survey
\citep[UKIDSS,][]{law07}, the Wide-field Infrared Survey Explorer
\citep[WISE,][]{wri10}, Pan-STARRS1 \citep{kai02,kai10},
and the Visible and Infrared Survey Telescope for Astronomy (VISTA)
Hemisphere Survey \citep[VHS,][]{mcm13}. In previous studies,
we have used those data to select candidate members of 
Ophiuchus and Upper Sco, many of which were then observed with spectroscopy 
\citep{esp18,luh18,esp20,luh20u}.
Meanwhile, we have identified candidates toward the Lupus clouds using data 
from the Gaia's second data release \citep{luh20lu} and have performed similar
analysis for all populations in Sco-Cen with the early installment of the
third data release of Gaia \citep[EDR3,][]{luh21}.
Other recent studies also have searched for members of Sco-Cen using 
deep optical and IR imaging \citep{lod18,lod21} and
Gaia data \citep{coo17,gol18,man18,ros18,can19,dam19,gal20b,tei20,gra21,ker21}.
In this work, we present spectroscopy of a few hundred candidate members of
Sco-Cen from our previous surveys with the goal of
improving the completeness of the spectroscopic samples in those regions.

\section{Spectroscopy of Candidates}

\subsection{Selection of Candidates}
\label{sec:select}

We have pursued spectroscopy of candidate members of Sco-Cen to measure their
spectral types and assess their youth. We selected the candidates
from our surveys of Sco-Cen \citep[e.g.,][]{esp20,luh20u,luh20lu,luh21},
giving the highest priority to candidates located within the Ophiuchus
field from \citet{esp18}, the triangular field from \citet{luh20u} that
encompasses the central concentration of stars in Upper Sco, and
the fields toward Lupus clouds 1--4 from \citet{luh20lu}.
We focused on candidates that lack spectroscopic data, but we included
stars that have previous classifications (primarily in Lupus) because
they were sufficiently bright for poor observing conditions.

\subsection{Observations}

We have obtained 293 spectra for 284 candidates in Sco-Cen.
Nine objects were observed with both optical and IR spectrographs.
The observations were performed with the 
4~m Blanco telescope at the Cerro Tololo Inter-American Observatory (CTIO),
the NASA Infrared Telescope Facility (IRTF), and the Gemini North telescope.
The instrument configurations are summarized in Table~\ref{tab:log}.
The COSMOS observations with the $0\farcs9$ and $1\farcs2$ slits were
performed on 2021 June 19--21 and 24/26, respectively.
We have made use of archival data for one additional object
that was observed through program 085.C-0887(A) (M. Fang) with the Infrared 
Spectrometer and Array Camera \citep{moo98} at the Unit Telescope 4
of the Very Large Telescope at the European Southern Observatory.
The configuration of that instrument produced a spectrum covering the
$H$ and $K$ bands with a resolution of $\sim$1000.
The 285 spectroscopic targets are listed in Table~\ref{tab:spec}, which
includes the instruments and dates for their observations.

\begin{figure}
\epsscale{1}
\plotone{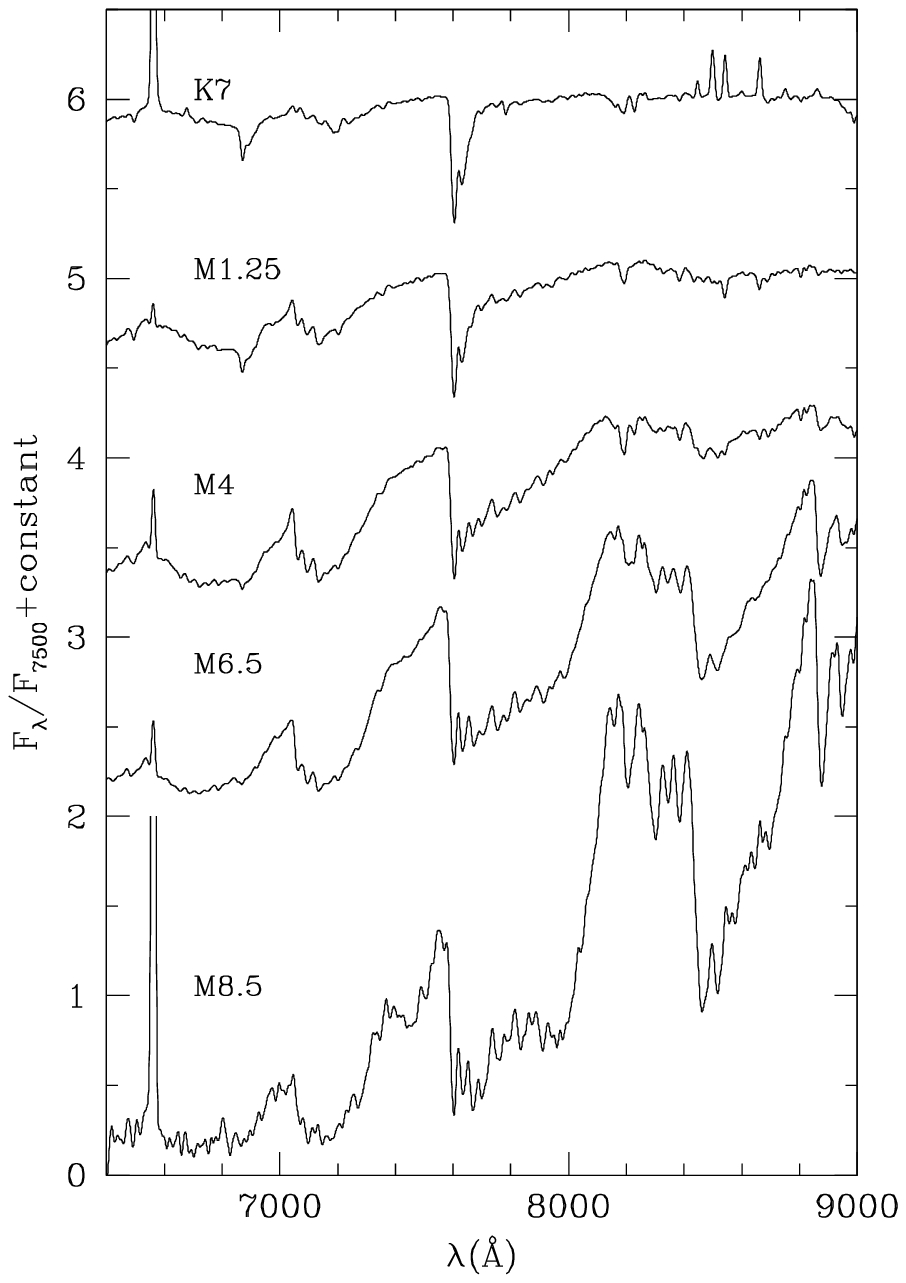}
\caption{
Examples of optical spectra of candidate members of Sco-Cen, which
are displayed at a resolution of 13~\AA.
The data used to create this figure are available.
}
\label{fig:op}
\end{figure}

\begin{figure}
\epsscale{1}
\plotone{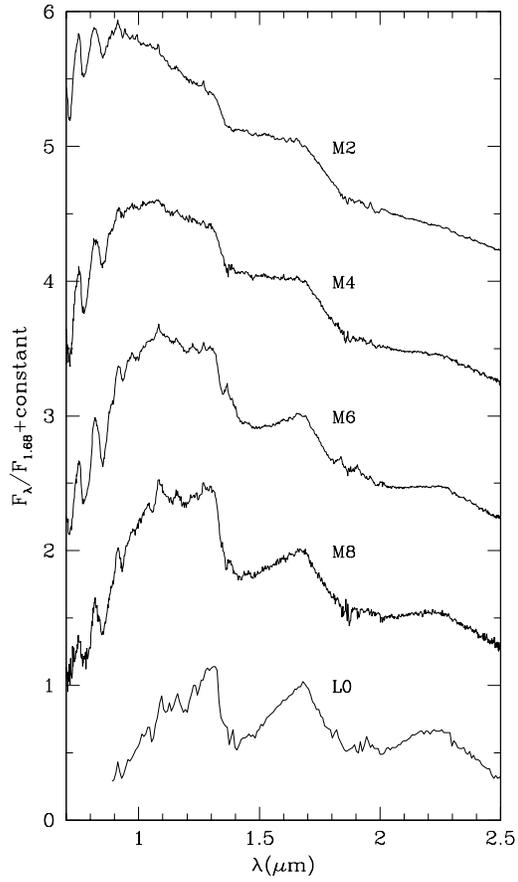}
\caption{
Examples of near-IR spectra of candidate members of Sco-Cen,
which are displayed at a resolution of $R=150$.
The data used to create this figure are available.
}
\label{fig:ir}
\end{figure}

We reduced the IRTF/SpeX data with the Spextool package \citep{cus04}, which
included correction of telluric absorption \citep{vac03}.
The CTIO/ARCoIRIS spectra were reduced with a modified version of Spextool. 
The spectra from the remaining instruments were reduced using routines
within IRAF. Examples of the reduced optical and IR
spectra are presented in Figures~\ref{fig:op} and \ref{fig:ir}, respectively.
The reduced spectra are available in electronic files associated with
those figures. Due to a malfunction, the telescope was not well-focused
during the COSMOS observations on June 19--21. As a result, the slit loss
varied with wavelength, and hence the slopes of those spectra are not reliable. 

\subsection{Spectral Classifications}

\begin{figure}
\epsscale{1}
\plotone{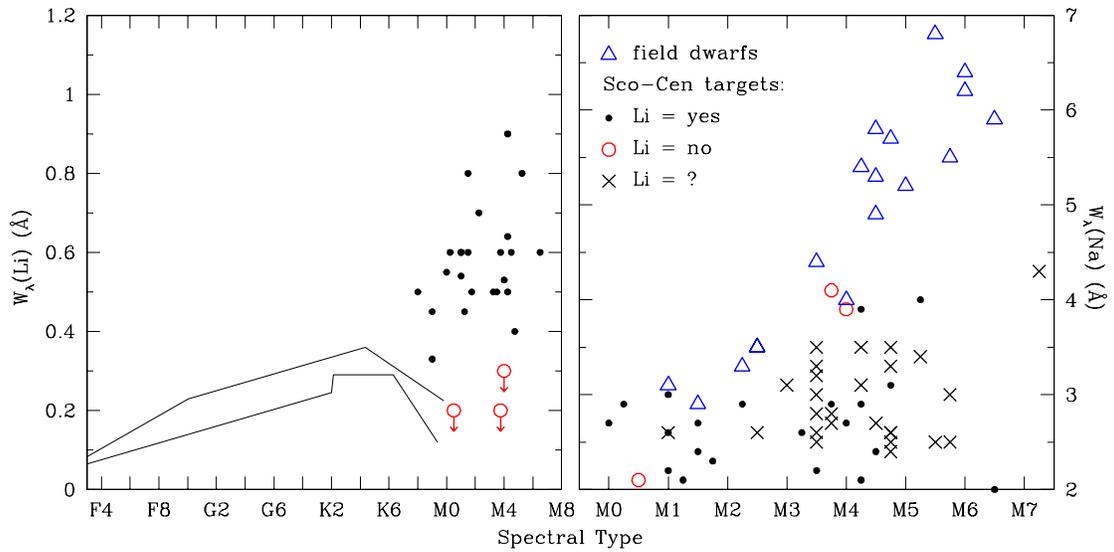}
\caption{
Equivalent widths of Li and Na versus spectral type for 
candidate members of Sco-Cen for which we have
obtained optical spectra (Table~\ref{tab:spec}).
The left diagram contains detections of Li and useful upper limits.
In the right diagram, we show Na measurements from all optical spectra,
including those that lack constraints on Li.
For comparison, we include the upper envelopes for Li data in IC~2602 (30~Myr)
and the Pleiades (125~Myr) \citep[upper and lower solid lines,][]{neu97}
in the left diagram and Na measurements for a sample of field dwarfs in the
right diagram.}
\label{fig:lina}
\end{figure}

We have analyzed the spectra with the same methods used in our previous
surveys in Sco-Cen and other star-forming regions \citep[e.g.,][]{esp18,luh18}.
In summary, we examined diagnostics of youth in the form of 
Li absorption at 6707~\AA\ and gravity-sensitive features like the Na doublet
near 8190~\AA\ and the near-IR H$_2$O absorption bands.
The use of Li and Na for constraining ages is illustrated in
Figure~\ref{fig:lina}, where we plot the equivalent widths of those
lines versus spectral type for the targets with optical spectra.
We have included the upper envelopes for Li data in IC~2602 (30~Myr) and the
Pleiades (125~Myr) from \citet{neu97} and Na data for a
sample of standard field dwarfs from our previous surveys and \citet{fil16}.
All of the Li detections are stronger than the upper envelope for
IC~2602, and thus are taken as evidence of youth.
Meanwhile, all of the stars with Na measurements but no useful constraints on
Li have sufficiently weak Na relative to the field dwarfs to indicate that they
are probably young.
We have measured spectral types through comparison to standard spectra
for field dwarfs \citep{kir91,kir97,hen94,cus05,ray09}
and members of star-forming regions \citep{luh97,luh99,luh17}.
We present our measurements of spectral types, equivalent widths of Li and Na,
and assessments of youth in Table~\ref{tab:spec}. The latter are based on
our spectra as well as age constraints available from previous work.
Most of the objects (269/285) have evidence of youth.

Table~\ref{tab:spec} contains a flag that indicates the populations
in Sco-Cen with which the kinematic data from Gaia EDR3 are consistent
based on the analysis in \citet{luh21} and a flag that indicates whether
each object is within the fields in Ophiuchus, Upper Sco, and Lupus
mentioned in Section~\ref{sec:select}. The kinematic flag is absent
for objects that lack parallax measurements with errors of $<1$~mas from
Gaia EDR3. Eight of the spectroscopic targets are flagged as non-members
for Sco-Cen based on data from Gaia EDR3.

\subsection{Classifications of Spectra from Previous Studies}

In addition to the spectra collected in this work, we have classified
IR spectra from previous studies for several candidates for 
substellar members of Ophiuchus and Upper Sco.
They consist of three sources toward Ophiuchus that were classified
as young by \citet{all20} and that we have not classified previously;
the two candidate members of Upper Sco that were observed spectroscopically
by \citet{lod21}; and three highly reddened objects toward Ophiuchus
that have been described as protostellar brown dwarfs by \citet{whe18}
and \citet{ria21} and that exhibit absorption features in spectra
from those studies.

\begin{figure}
\epsscale{1}
\plotone{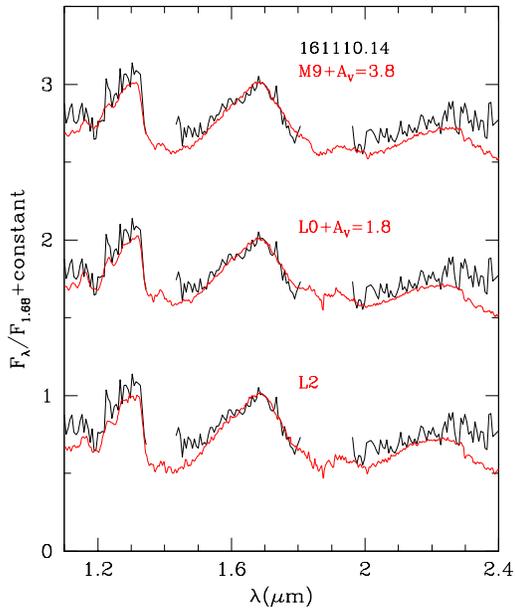}
\caption{
Our reduction of an IR spectrum from \citet{lod21} for a candidate
late-type member of Upper Sco, UGCS~J161110.14$-$214516.8,
which has been binned to a resolution
of R=150. For comparison, we have included the young standard spectra
from \citet{luh17} for M9, L0, and L2. Reddening has been applied to
the M9 and L0 spectra to match the slope of the Upper Sco candidate.
}
\label{fig:lod}
\end{figure}

We have classified the spectra from \citet{all20} and \citet{lod21}
through comparison to the young standard spectra from \citet{luh17}.
The resulting spectral types are similar to those from \citet{all20}
and are earlier than those from \citet{lod21}.
In Figure~\ref{fig:lod}, we present the spectrum of one of the two
objects from \citet{lod21}, UGCS J161110.14$-$214516.8, and we
include comparisons to standard spectra for M9, L0, and L2.
The M9 and L0 spectra have been reddened to match the slope of 
UGCS J161110.14$-$214516.8 while no reddening is needed for the L2 spectrum. 
The three standards in Figure~\ref{fig:lod} provide similar fits,
which illustrates the degeneracy between spectral type and reddening
for young L dwarfs \citep{luh17}. \citet{lod21} classified 
UGCS J161110.14$-$214516.8 as L6, but the standards later than L2 from
\citet{luh17} are too red to match its slope.

\begin{figure}
\epsscale{1}
\plotone{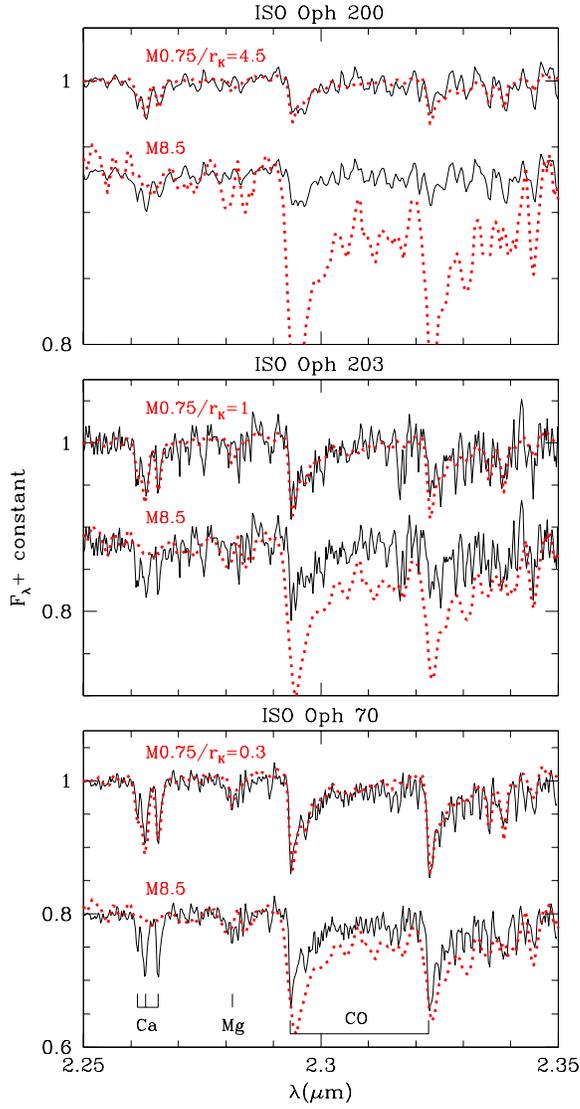}
\caption{
IR spectra from \citet{whe18} and \citet{ria21} for three candidate protostellar
brown dwarfs in Ophiuchus, each of which is compared to 1) a young early-M star
that has been artificially veiled to match the strengths of the Ca and CO
features \citep{cov10} and 2) a young late-M object \citep{all13a}.
}
\label{fig:iso}
\end{figure}

The spectra of ISO~Oph~70, 200, and 203 from \citet{whe18} and 
\citet{ria21} contain absorption features that can be used to constrain 
the spectral types of the stars, which was not attempted in those studies.
Among the detected absorption lines, the most useful ones for spectral 
classification are produced by Ca, Mg, and CO in the $K$ band. 
In Figure~\ref{fig:iso}, we show the spectra for a wavelength range 
encompassing those features. To facilitate comparison to standard spectra, 
the spectra have been divided by low-order fits to the continua.
The sources exhibit large IR excesses \citep{bon01,dun08}, so the observed 
$K$-band spectra may include continuous emission from circumstellar dust, which 
would dilute the absorption lines from the stellar photospheres.
The lines in Figure~\ref{fig:iso} span a small enough range of wavelengths
that they should be subjected to the same degree of continuum veiling,
in which case the relative line strengths will be independent of veiling.
We have estimated spectral types from those relative strengths through
comparison to spectra of young stars that have optical spectral types 
\citep[e.g.,][]{cov10,all13a}. 
For all three objects, we find that the lines are best fit by spectral
types of early M. To illustrate these fits, we have included
in Figure~\ref{fig:iso} comparisons to a spectrum of the young
star TWA~25 \citep[M0.75,][]{zuc04,luh17} after the application of continuum
veiling, which is quantified as
$r_{K}=I_{2.2}({\rm IR~excess})/I_{2.2}({\rm star})$.
In addition, we show comparisons to the young brown dwarf TWA~26
\citep[M8.5,][]{giz02,luh17}, which differs significantly from the
Ophiuchus sources in terms of the relative strengths of Ca and CO.
The Ophiuchus objects are likely to be younger than the association
containing TWA~25 and TWA~26 \citep[$\sim10$~Myr,][]{web99,mam05},
in which case they would have lower surface gravities. 
However, late-M standards with lower gravities would produce worse fits
since they would have even stronger CO bands.
In our low-resolution IR spectrum of ISO~Oph~200, the CO bands are in
emission rather than absorption, indicating the presence of strongly variable
emission in that feature from circumstellar material. 
Indeed, in the spectrum of that object in Figure~\ref{fig:iso}, the first
bandhead of absorption at 2.29~\micron\ appears to contain weak emission
within it. Accounting for such line veiling of CO could improve the match
to a late-M type, but the relative strengths of Ca and the other features
at $<2.29$~\micron\ would remain discrepant.

The spectral types of early M that we derive for ISO~Oph~70, 200, and 203 
should correspond to masses in the vicinity of 0.6~$M_\odot$ based
on the temperatures predicted for ages of $\lesssim10$~Myr by
theoretical evolutionary models \citep{bar98,bar15}.
Young brown dwarfs are expected to have spectral types of $\gtrsim$M6.5,
which are inconsistent with the spectra of those sources.
The substellar masses previously reported for ISO~Oph~70, 200, and 203
have been derived from estimates of their bolometric luminosities using
predicted relations between luminosity and mass for protostars \citep{ria21}.
However, the luminosities of protostars can be underestimated if they
are detected primarily in scattered light at the near-IR wavelengths
where photospheric fluxes are measured. In addition, the predicted
relations between luminosity and mass for protostars are uncertain
\citep{dun14}.

The spectral types that we have measured in this section are included
in Table~\ref{tab:usco}, where we compile spectral types from this work
and previous studies for all objects toward Upper Sco and Upper Sco that
have spectral classifications and evidence of youth.

\section{Status of Spectroscopic Samples in Sco-Cen}

\subsection{Compilations of Spectroscopically Classified Stars}

Recent studies have presented compilations of candidate members
of Ophiuchus, Upper Sco, and Lupus that have been observed spectroscopically
and that exhibit evidence of youth \citep{esp20,luh20u,luh20lu}.
In addition, \citet{luh21} has performed a census of
all of the Sco-Cen populations using Gaia EDR3 and has compiled
the spectral types that are available (including those from this work)
for the resulting candidates.
In Tables~\ref{tab:usco} and \ref{tab:lup}, we present our latest
compilations of candidate members of Ophiuchus/Upper Sco and Lupus,
respectively, that have spectral classifications and that satisfy the following
additional criteria. We have only considered Ophiuchus and Upper Sco 
candidates that are within the boundary of Upper Sco from \citet{dez99}
($l=343$--$360\arcdeg$, $b=10$--$30\arcdeg$) and
Lupus candidates that are within the fields toward clouds 1--4 from 
\citet{luh20lu}. The Ophiuchus/Upper Sco and Lupus catalogs contain objects
that have spectroscopy and that 
1) lack Gaia EDR3 astrometry and have evidence of youth
or 2) have Gaia EDR3 parallaxes and proper motions that satisfy
the criteria for membership in Ophiuchus/Upper Sco or Lupus \citep{luh21} and 
that have evidence of youth or lack constraints on youth. One consequence of 
these criteria is that they exclude young stars projected against Ophiuchus,
Upper Sco, and Lupus that have astrometry indicative of other populations in 
Sco-Cen. For close pairs that are resolved by Gaia but were likely unresolved
during the spectroscopy, we have listed only the components that are brighter
in the $G$ band from Gaia. The catalog of Sco-Cen candidates in \citet{luh21} 
includes the fainter components of such pairs if they satisfy the kinematic
criteria for membership. The contents of Tables~\ref{tab:usco} and 
\ref{tab:lup} consist of source names from
Gaia EDR3, UKIDSS (for Ophiuchus Upper Sco), 2MASS, and various
other catalogs, equatorial coordinates, available spectral types,
our adopted classifications, and a flag indicating the populations
with which the kinematic data from Gaia EDR3 are consistent.
In Table~\ref{tab:usco}, we also include a flag indicating whether each
object is located in the Ophiuchus field from \citet{esp18} or the triangular
Upper Sco field from \citet{luh20u}. The catalogs for Ophiuchus, Upper Sco, and
Lupus contain 419, 1723, and 132 sources, respectively, 61, 237, and 20 of
which are later than M6.

\subsection{Completeness of Spectroscopic Samples}

\begin{figure}
\epsscale{1}
\plotone{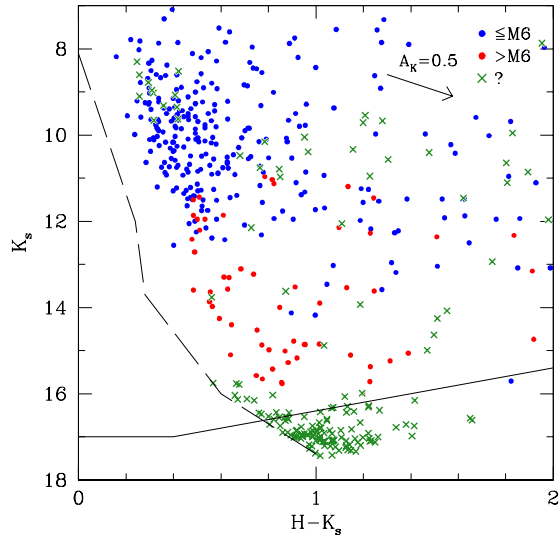}
\caption{
Near-IR color-magnitude diagram from 2MASS, UKIDSS, and VHS
for candidate members of Ophiuchus that are located within the field 
from \citet{esp18} and that have spectral classifications 
(red and blue points, Table~\ref{tab:usco}).
We also show the remaining IR sources in that field that are not rejected by
available membership constraints (green crosses).
We have marked a boundary that follows the lower envelope of the known members
of Upper Sco and Ophiuchus (dashed line) and the UKIDSS completeness limit
(solid line).
}
\label{fig:hk2}
\end{figure}

\begin{figure}
\epsscale{1}
\plotone{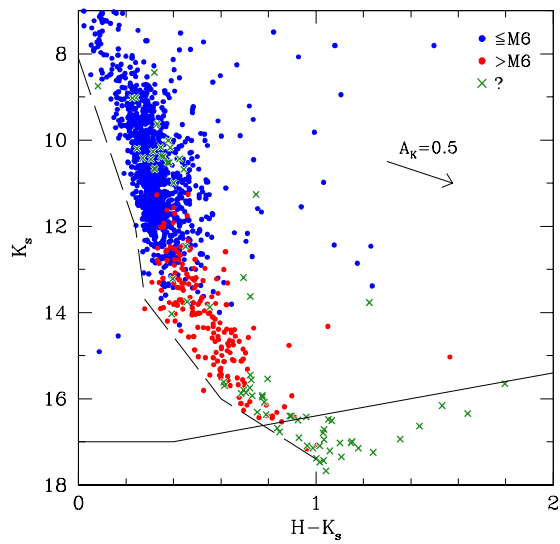}
\caption{
Same as Figure~\ref{fig:hk2}, but for candidate members of Upper Sco 
that are located within the triangular field from \citet{luh20u},
excluding Ophiuchus.
}
\label{fig:hk3}
\end{figure}

\begin{figure}
\epsscale{1}
\plotone{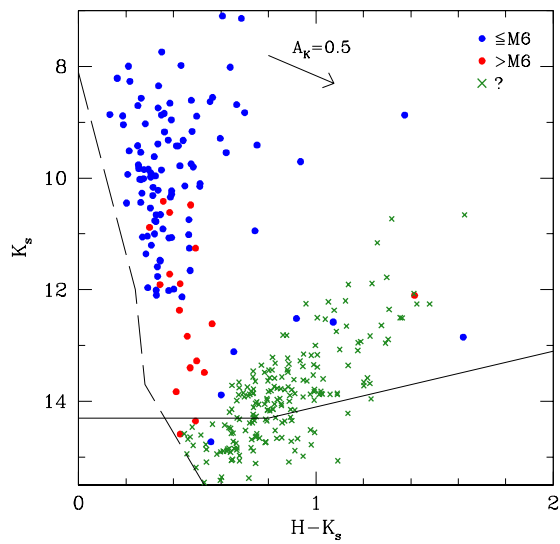}
\caption{
Same as Figure~\ref{fig:hk2}, but for candidate members of Lupus 
(Table~\ref{tab:lup}) that are located within the fields toward 
clouds 1--4 from \citet{luh20lu}. These data are from 2MASS.
}
\label{fig:hk1}
\end{figure}

We wish to characterize the completeness of the spectroscopic samples
in the Ophiuchus field from \citet{esp18}, the triangular field in
Upper Sco from \citet{luh20u}, and the fields toward Lupus clouds 1--4 from
\citet{luh20lu}. To do that, we employ color-magnitude diagrams (CMDs) 
constructed from $H$ and $K_s$ because they
provide the best sensitivity to members of these regions given the
typical colors of young stars and the depths of the available imaging surveys.
We use photometry from 2MASS, UKIDSS, and VHS for Ophiuchus and Upper Sco
and photometry from 2MASS for Lupus.  We present diagrams of $K_s$ versus 
$H-K_s$ for the fields in Ophiuchus, Upper Sco, and Lupus in 
Figures~\ref{fig:hk2}, \ref{fig:hk3}, and \ref{fig:hk1}, respectively.
For the Ophiuchus and Upper Sco CMDs, we have marked the completeness
limit estimated for the UKIDSS data by \citet{luh20u}. For the Lupus CMD,
we show the completeness limit for 2MASS from \citet{skr06}.
The CMDs contain the sources with spectral classifications from
Tables~\ref{tab:usco} and \ref{tab:lup} that are located within the fields
in question, which are plotted with blue and red points for spectral
types of $\leq$M6 and $>$M6, respectively.
We also have included all other sources in the fields with the exception of 
those that 1) have data indicating that they are not young, 2) have kinematics
from Gaia EDR3 that are inconsistent with membership \citep{luh21}, 
3) are resolved as galaxies in available imaging,
4) are rejected by CMDs from the imaging surveys mentioned in
Section~\ref{sec:intro} \citep{luh18,esp20,luh20u}, or 5) appear below
the dashed line in the CMDs in Figures~\ref{fig:hk2}--\ref{fig:hk1},
which was selected by \citet{luh18} to follow the lower envelope of the
sequence of known members of Upper Sco and Ophiuchus.

In the CMD for Ophiuchus, the number of sources that lack spectra
is fairly small relative to the spectroscopic sample
for a wide range of magnitudes and reddenings.
The CMD indicates that the spectroscopic sample has a high level of 
completeness for an extinction-corrected magnitude of $K<15.5$ for $A_K<0.8$. 
There are 44 objects that lack spectra at $H-K_s<2$ and $K_s<15$,
18 of which have Gaia data that support membership, making them strong
candidates. All of those Gaia candidates have $H-K_s<1.3$ and $K_s\lesssim12$.
Many of the redder objects without spectral classifications are also 
probable members based on IR excesses and other signatures of youth 
(e.g., outflows).
The CMD for Upper Sco contains 32 stars with undetermined membership at
$K\lesssim14$, half of which are candidates based on Gaia data.
The other half of those stars are bright enough for Gaia detections
but lack measurements of parallaxes and proper motions.
The completeness for Upper Sco is high at $K_s\lesssim15.5$.
Between $K_s=15.5$ and the magnitude at which the UKIDSS completeness limit
intersects with the Upper Sco sequence ($K\sim16.5$), the numbers of
objects with spectra and without spectra are roughly similar.
Finally, the Lupus CMD suggests that the spectroscopic sample
is complete down $K_s\sim14$ for low extinctions ($A_K<0.2$). 
For the range of extinctions exhibited by most members ($A_K<0.5$), 
the sample should be mostly complete for an extinction-corrected magnitude 
of $K_s\sim13$.

Assuming the median distances of the populations and the
$K$-band bolometric correction for young late-type objects \citep{fil15},
the completeness limits of $K_s=15.5$, 15.5, and 13 for Ophiuchus, Upper Sco,
and Lupus from the preceding analysis correspond to masses of $\sim$0.008, 
0.014, and 0.035~$M_\odot$ for ages of 3, 10, and 5~Myr, respectively
\citep{cha00,bar15}.

\subsection{Prospects for Future Spectroscopy in Sco-Cen}

The completeness of the spectroscopic samples in Ophiuchus and the center
of Upper Sco can be improved by obtaining spectra of the remaining candidates
at $K_s<15$ discussed in the previous section.
The completeness limits of those samples and the sample in Lupus can be
extended to fainter magnitudes through deeper optical and IR imaging
and spectroscopy of the resulting candidates.
Outside of the fields in Ophiuchus, Upper Sco, and Lupus in which we
have examined completeness, several thousand candidate members of Sco-Cen
have been identified using data from Gaia EDR3 \citep{luh21}, most
of which lack spectral classifications. The most appealing subsets for
spectroscopy include the hundreds of candidates that have IR excesses from disks
\citep{luh21b}, particularly in the oldest populations of V1062~Sco and 
UCL/LCC, and the hundreds of candidates with colors indicative of 
brown dwarfs ($\gtrsim$M6.5).

\section{Conclusions}

We have presented optical and IR spectroscopy of 285 candidate members
of the Sco-Cen complex that have been identified in our previous studies.
The results are summarized as follows:

\begin{enumerate}

\item
We have measured spectral types for the Sco-Cen candidates and have
assessed their youth using spectral diagnostics and other available data.
Evidence of youth is found for 269 of the 285 candidates.
In addition, we have compiled all objects with spectral classifications 
and evidence of youth in Ophiuchus, Upper Sco, and the fields toward 
Lupus clouds 1--4 from \citet{luh20lu}.  The resulting catalogs 
contain 419, 1723, and 132 sources for the three regions, respectively, 
61, 237, and 20 of which are later than M6.

\item
We have classified IR spectra from previous studies for several
candidates for substellar members of Ophiuchus and Upper Sco. 
In the most notable result, we find that three sources in Ophiuchus 
that have been proposed to be protostellar brown dwarfs 
(ISO~Oph~70, 200, 203) have spectral types of early M, which are
earlier than expected for young brown dwarfs based on
evolutionary models ($\gtrsim$M6.5) and instead are indicative of
stellar masses ($\sim0.6$~$M_\odot$).

\item
We have used near-IR CMDs to characterize the completeness of the
spectroscopic samples in the Ophiuchus field from \citet{esp18},
the triangular field in Upper Sco from \citet{luh20u}, and the fields
toward Lupus clouds 1--4 from \citet{luh20lu}.
Those samples have high levels of completeness at $K_s\lesssim15.5$, 15.5, 
and 13 for Ophiuchus, Upper Sco, and Lupus, respectively, for the ranges of
extinctions encompassing most members. Those magnitudes
correspond to masses of $\sim$0.008, 0.014, and 0.035~$M_\odot$ for 
ages of 3, 10, and 5~Myr, respectively, based on evolutionary models
\citep{cha00,bar15}.

\end{enumerate}

\acknowledgements

K. L. acknowledges support from NASA grant 80NSSC18K0444 for portions of
this work.  We thank Katelyn Allers for providing the modified version of
Spextool for use with ARCoIRIS data.
The IRTF is operated by the University of Hawaii under contract 80HQTR19D0030
with NASA. The data at CTIO were obtained through programs
2016A-0157, 2021A-0012, and 2021A-0015 at NOIRLab.
CTIO and NOIRLab are operated by the Association of Universities for Research in
Astronomy under a cooperative agreement with the NSF. The Gemini data were
obtained through program GN-2020A-Q-218 (2020A-0066).
Gemini Observatory is operated by AURA under a cooperative agreement with
the NSF on behalf of the Gemini partnership: the NSF (United States), the NRC
(Canada), CONICYT (Chile), Minist\'{e}rio da Ci\^{e}ncia, 
Tecnologia e Inova\c{c}\~{a}o (Brazil), Ministerio de Ciencia, 
Tecnolog\'{i}a e Innovaci\'{o}n Productiva (Argentina), and
Korea Astronomy and Space Science Institute (Republic of Korea).
This work used data from the European Space Agency (ESA)
mission Gaia (\url{https://www.cosmos.esa.int/gaia}), processed by
the Gaia Data Processing and Analysis Consortium (DPAC,
\url{https://www.cosmos.esa.int/web/gaia/dpac/consortium}). Funding
for the DPAC has been provided by national institutions, in particular
the institutions participating in the Gaia Multilateral Agreement.
2MASS is a joint project of the University of Massachusetts and IPAC
at Caltech, funded by NASA and the NSF.
The Center for Exoplanets and Habitable Worlds is supported by the
Pennsylvania State University, the Eberly College of Science, and the
Pennsylvania Space Grant Consortium.

\clearpage

\begin{deluxetable}{llll}
\tabletypesize{\scriptsize}
\tablewidth{0pt}
\tablecaption{Observing Log\label{tab:log}}
\tablehead{
\colhead{Telescope/Instrument\tablenotemark{a}} &
\colhead{Disperser/Aperture} &
\colhead{Wavelengths/Resolution} &
\colhead{Targets}}
\startdata
CTIO 4~m/COSMOS & red VPH/$0\farcs9$ slit & 0.55--0.95~\micron/3~\AA & 19 \\
CTIO 4~m/COSMOS & red VPH/$1\farcs2$ slit & 0.55--0.95~\micron/4~\AA & 36 \\
CTIO 4~m/ARCoIRIS & 110.5 l~mm$^{-1}$ + prism/$1\farcs1$ slit &0.8--2.47~\micron/R=3500 & 8 \\
Gemini North/GNIRS & 31.7~l~mm$^{-1}$/$1\arcsec$ slit & 0.9--2.5~\micron/R=600 & 14 \\
IRTF/SpeX & prism/$0\farcs8$ slit & 0.8--2.5~\micron/R=150 & 216
\enddata
\tablenotetext{a}{
The Gemini Near-Infrared Spectrograph (GNIRS) and SpeX are described by
\citet{eli06} and \citet{ray03}, respectively.
The Cerro Tololo Ohio State Multi-Object Spectrograph (COSMOS) is
based on an instrument described by \citet{mar11}.}
\end{deluxetable}

\begin{deluxetable}{ll}
\tabletypesize{\scriptsize}
\tablewidth{0pt}
\tablecaption{Spectroscopic Data for Candidate Members of Sco-Cen\label{tab:spec}}
\tablehead{
\colhead{Column Label} &
\colhead{Description}}
\startdata
Gaia & Gaia EDR3 source name \\
UGCS & UKIDSS Galactic Clusters Survey source name \\
2MASS & 2MASS Point Source Catalog source name \\
Name & Other source name \\
RAdeg & Right ascension (ICRS) \\
DEdeg & Declination (ICRS) \\
Ref-Pos & Reference for right ascension and declination\tablenotemark{a} \\
SpType & Spectral type from this work\tablenotemark{b}\\
f\_EWLi & Flag for EWLi\\
EWLi & Equivalent width of Li\tablenotemark{c}\\
EWNa & Equivalent width of Na\tablenotemark{c}\\
Instrument & Instrument used for spectral classification\\
Date & Date of spectroscopy\\
Young & Is the candidate young based on spectroscopy or other diagnostics?\\
Kin & Kinematic population\tablenotemark{d}\\
Pos & Position in Sco-Cen\tablenotemark{e}
\enddata
\tablenotetext{a}{Sources of the right ascension and declination are Gaia EDR3
(Epoch 2016.0), DR10 of the UKIDSS Galactic Clusters Survey,
and the 2MASS Point Source Catalog.}
\tablenotetext{b}{Uncertainties are 0.25 and 0.5~subclass for optical and
IR spectral types, respectively, unless indicated otherwise.}
\tablenotetext{c}{Typical uncertainties are 0.05 and 0.3~\AA\ for
Li and Na, respectively.}
\tablenotetext{d}{Parallax and proper motion offset from Gaia EDR3 are 
consistent with membership in these Sco-Cen populations based on the criteria
in \citet{luh21}: u = Upper Sco; o = Ophiuchus; l = Lupus; v = V1062 Sco;
c = UCL/LCC; n = none of these populations.}
\tablenotetext{e}{Celestial coordinates within these regions:
o = the Ophiuchus field from \citet{esp18};
u = the triangular field in Upper Sco from \citet{luh20u}, excluding Ophiuchus;
l = the fields encompassing Lupus clouds 1--4 from \citet{luh20lu}.}
\tablecomments{
The table is available in its entirety in machine-readable form.}
\end{deluxetable}

\begin{deluxetable}{ll}
\tabletypesize{\scriptsize}
\tablewidth{0pt}
\tablecaption{Candidate Members of Ophiuchus and Upper Sco with Spectral Classifications at $l=343$--$360\arcdeg$ and $b=10$--$30\arcdeg$\label{tab:usco}}
\tablehead{
\colhead{Column Label} &
\colhead{Description}}
\startdata
Gaia & Gaia EDR3 source name \\
UGCS & UKIDSS Galactic Clusters Survey source name \\
2MASS & 2MASS Point Source Catalog source name \\
Name & Other source name \\
RAdeg & Right ascension (ICRS) \\
DEdeg & Declination (ICRS) \\
Ref-Pos & Reference for right ascension and declination\tablenotemark{a} \\
SpType & Spectral type \\
r\_SpType & Spectral type reference\tablenotemark{b} \\
Adopt & Adopted spectral type \\
Kin & Kinematic population\tablenotemark{c} \\
Pos & Position in Sco-Cen\tablenotemark{d}
\enddata
\tablenotetext{a}{Sources of the right ascension and declination are Gaia EDR3
(Epoch 2016.0), DR6 of VISTA VHS, DR10 of the UKIDSS Galactic Clusters Survey, 
the 2MASS Point Source Catalog, and high-resolution
imaging \citep{ire11,kra14,lac15,bry16}.}
\tablenotetext{b}{
(1) \citet{hou88};
(2) \citet{can93};
(3) \citet{luh18};
(4) \citet{daw14};
(5) \citet{esp18};
(6) \citet{luh20u};
(7) \citet{riz15};
(8) \citet{pre98};
(9) \citet{lod08};
(10) \citet{bon14};
(11) this work;
(12) \citet{lod06};
(13) \citet{kun99};
(14) \citet{hil69};
(15) \citet{pec16};
(16) \citet{pen16};
(17) \citet{hou82};
(18) \citet{cor84};
(19) \citet{all13b};
(20) \citet{mar04};
(21) \citet{mar10};
(22) \citet{sle08};
(23) \citet{chi20};
(24) \citet{pre02};
(25) \citet{mor01};
(26) \citet{vie03};
(27) \citet{rei08};
(28) \citet{ard00};
(29) \citet{wal94};
(30) \citet{kir10};
(31) \citet{all13a};
(32) \citet{fah16};
(33) \citet{tor06};
(34) \citet{giz02};
(35) \citet{her14};
(36) \citet{sle06};
(37) \citet{ria06};
(38) \citet{bes17};
(39) \citet{lod18};
(40) \citet{kra15};
(41) \citet{cod17};
(42) \citet{laf11};
(43) \citet{lac15};
(44) \citet{dav19b};
(45) \citet{ans16};
(46) \citet{pra02};
(47) \citet{pre01};
(48) \citet{cru03};
(49) \citet{bej08};
(50) \citet{her09};
(51) \citet{lod21};
(52) measured in this work with the most recently published spectrum;
(53) \citet{kra07};
(54) \citet{lod11};
(55) \citet{mue11};
(56) \citet{kra09};
(57) \citet{pec12};
(58) \citet{bil11};
(59) \citet{laf08};
(60) \citet{luh17};
(61) \citet{man16};
(62) \citet{dav16b};
(63) \citet{sta17};
(64) \citet{sta18};
(65) \citet{coh79};
(66) \citet{pra03};
(67) \citet{eis05};
(68) \citet{mur69};
(69) \citet{mar98a};
(70) \citet{lev75};
(71) \citet{pra07};
(72) \citet{mcc10};
(73) \citet{lod15};
(74) \citet{dav16a};
(75) \citet{car06};
(76) \citet{luh05usco};
(77) \citet{mar98b};
(78) \citet{esp20};
(79) \citet{all20};
(80) \citet{cie10};
(81) \citet{bow11};
(82) \citet{bow14};
(83) \citet{rom12};
(84) \citet{all07};
(85) \citet{mer10};
(86) \citet{eri11};
(87) \citet{wil05};
(88) \citet{bra97};
(89) \citet{mah03};
(90) \citet{alv12};
(91) \citet{gre95};
(92) \citet{luh99};
(93) \citet{bou92};
(94) \citet{alv10};
(95) \citet{str49};
(96) \citet{muz12};
(97) \citet{wil99};
(98) \citet{nat02};
(99) \citet{man15};
(100) \citet{tes02};
(101) \citet{com10};
(102) \citet{cus00};
(103) \citet{bow17};
(104) \citet{luh97};
(105) \citet{gee11};
(106) \citet{gat06};
(107) \citet{gee07};
(108) \citet{pat93};
(109) \citet{vrb93};
(110) \citet{ryd80}.}
\tablenotetext{c}{Parallax and proper motion offset from Gaia EDR3 are 
consistent with membership in these Sco-Cen populations based on the criteria
in \citet{luh21}: u = Upper Sco; o = Ophiuchus; l = Lupus; v = V1062 Sco;
c = UCL/LCC.}
\tablenotetext{d}{Celestial coordinates within these regions:
o = the Ophiuchus field from \citet{esp18};
u = the triangular field in Upper Sco from \citet{luh20u}, excluding Ophiuchus.}
\tablecomments{
The table is available in its entirety in machine-readable form.}
\end{deluxetable}

\begin{deluxetable}{ll}
\tabletypesize{\scriptsize}
\tablewidth{0pt}
\tablecaption{Candidate Members of Lupus Clouds 1--4 with Spectral Classifications\label{tab:lup}}
\tablehead{
\colhead{Column Label} &
\colhead{Description}}
\startdata
Gaia & Gaia EDR3 source name \\
2MASS & 2MASS Point Source Catalog source name \\
Name & Other source name \\
RAdeg & Right ascension (ICRS)\tablenotemark{a} \\
DEdeg & Declination (ICRS)\tablenotemark{a} \\
SpType & Spectral type \\
r\_SpType & Spectral type reference\tablenotemark{b} \\
Adopt & Adopted spectral type \\
Kin & Kinematic population\tablenotemark{c}
\enddata
\tablenotetext{a}{Coordinates are from Gaia EDR3 (Epoch 2016.0) when
available and otherwise are from the 2MASS Point Source Catalog.}
\tablenotetext{b}{
(1) this work;
(2) \citet{app83};
(3) \citet{hey89};
(4) \citet{hug94};
(5) \citet{kun99};
(6) \citet{her14};
(7) \citet{alc17};
(8) \citet{alc14};
(9) \citet{kra97};
(10) \citet{tor06};
(11) \citet{her77};
(12) \citet{mar94};
(13) \citet{pec16};
(14) \citet{mor11};
(15) \citet{rom12};
(16) \citet{all07};
(17) \citet{com13};
(18) \citet{muz14};
(19) \citet{com03};
(20) \citet{man13};
(21) \citet{muz15};
(22) \citet{hou82};
(23) \citet{blo06}.}
\tablenotetext{c}{Parallax and proper motion offset from Gaia EDR3 are 
consistent with membership in these Sco-Cen populations based on the criteria
in \citet{luh21}: u = Upper Sco; l = Lupus; c = UCL/LCC.}
\tablecomments{
The table is available in its entirety in machine-readable form.}
\end{deluxetable}

\clearpage

\clearpage

\end{document}